\documentclass[fleqn,11pt]{wlscirep}
\usepackage[utf8]{inputenc}
\usepackage[T1]{fontenc}

\usepackage{bm}
\usepackage{lineno}
\title{Topological superfluids in two-dimensional Fermi gas with Rashba spin-orbit coupling}
\author[1]{Xiaosen Yang}
\author[2]{Ho-Kin Tang}
\author[2]{Noah Fan Qi Yuan}
\author[3]{Beibing Huang}
\author[4]{Guangcan Guo}
\author[4,*]{Ming Gong}
\author[2,5,$\dag$]{Hai-Qing Lin} 
\affil[1]{Department of Physics, Jiangsu University, Zhenjiang, 212013, P. R. China}
\affil[2]{Shenzhen JL Computational Science And Applied Research Institute, Shenzhen, 518131, P. R. China}
\affil[3]{Department of Physics, Yancheng Institute of Technology, Yancheng, 224051, P. R. China}
\affil[4]{CAS Key Lab of Quantum Information, University of Science and Technology of China, Hefei, 230026, P.R. China and Synergetic Innovation Center of Quantum Information and Quantum Physics, University of Science and Technology of China, Hefei, 230026, P.R. China}
\affil[5]{Beijing Computational Science Research Center, Beijing, 100084, P. R. China}
\affil[*]{gongm@ustc.edu.cn}
\affil[$\dag$]{haiqing0@csrc.ac.cn}

\begin{abstract}
The realization of spin-orbit coupling~(SOC) in ultracold atoms has triggered an intensive exploring of topological superfluids in the degenerate Fermi gases based on mean-field theory, which has not yet been reported in experiments. Here, we demonstrate the topological phase transitions in the system via the numerically exact quantum Monte Carlo method. Without prior assumptions, our unbiased real-space calculation shows that spin-orbit coupling can stabilize an unconventional pairing in the weak SOC regime, in which the Fulde-Ferrell-Larkin-Ovchinnikov pairing coexists with the Bardeen-Cooper-Schrieffer pairing. Furthermore, we use the jumps in the spin polarization at the time-reversal invariant momenta to qualify the topological phase transition, where we find the critical exponent deviated from the mean-field theory. Our results pave the way for the searching of unconventional pairing and topological superfluids with degenerate Fermi gases.
\end{abstract}
\begin{document}

\flushbottom
\maketitle
\section*{Introduction}

\begin{figure*}[hbt!]
\centering
\includegraphics[width=1\textwidth]{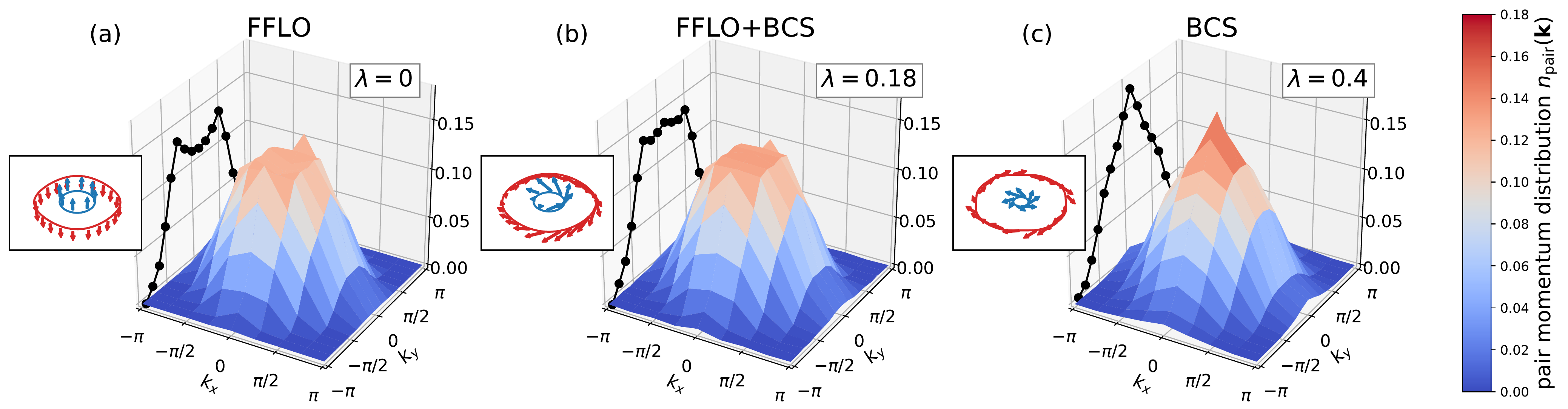}
\caption{\textbf{Competition between the Fulde-Ferrell-Larkin-Ovchinnikov~(FFLO) pairing and the Bardeen-Cooper-Schrieffer~(BCS) pairing}: The figure shows pair momentum distribution $n_{pair}(\textbf{k})$ for the spin-orbit coupling $\lambda=0,0.18,0.4$ (from left to right) with attractive interaction $U=3.5$, Zeeman field $h=1.7$, inverse temperature $\beta=10$ and filling factor $n = 0.375$. The FFLO-type pairing with nonzero momentum peaks at pair momentum distribution while the BCS-type pairing with the peak at the zero momentum. The FFLO+BCS pairing has both the zero momentum and nonzero momentum peaks at pair momentum distribution. With increasing spin-orbit coupling, the type of pairing goes from FFLO, FFLO+BCS, to BCS. In the inset, we illustrate the Fermi surface for the pairing in different phases with the arrows pointing to the direction of $\langle S_k \rangle$. We simulate the system in a $8 \times 16$ periodic supercell.} \label{fig.1}
\end{figure*}

Spin-orbit coupling~(SOC) has triggered extensive concerns in condensed matters and ultracold atoms for its essential role in many exotic phenomena, especially for the realization of topological phases such as topological insulator~\cite{RevModPhys.82.3045,qiRevModPhys2011}, topological superconductors~\cite{qiRevModPhys2011,Sato_2017}, spin Hall effects~\cite{qixiaoliangPhysRevB2006,PhysRevLett.97.240401}, Weyl semimetal~\cite{BurkovPhysRevLett2011, wenhongmingPhysRevX2015, ArmitageRevModPhys2018}, Floquet topological phase~\cite{lindnernaturephy2011,rudnerPhysRevX2013, xuscience2015} and so on. In ultracold atoms, SOC can be realized both in fermionic~\cite{PhysRevLett.109.095301, PhysRevLett.109.095302} and bosonic~\cite{PhysRevLett.102.130401, naturelin2009synthetic} systems by Raman couplings. These milestone breakthroughs have shed the light on exploring various novel phases with different topology and pairings in experiments \cite{PhysRevLett.97.240401,PhysRevLett.103.035301,PhysRevA.84.031608,PhysRevA.87.063610,banerjeenaturephy2013ferromagnetic, bertnaturephysics2011direct, PhysRevLett.109.085302,PhysRevLett.108.225301,PhysRevLett.105.160403,PhysRevLett.107.195305,reviewHzhai}.

The ultracold atoms also provide a clean and ideal platform to study the unconventional pairings and related phenomena in the Fermi gases~\cite{Mitra2018}.  In solid materials, the Fermi energy is much larger than the Zeeman splitting and SOC, thus only the Bardeen-Cooper-Schrieffer~(BCS) pairing is favored in most of the materials. As a result, while the Fulde-Ferrell-Larkin-Ovchinnikov~(FFLO)~\cite{FFstate,LOstate} superconductors with finite momentum pairings have been predicted through the competition between magnetism and pairing energies in the 1960s, it is hard to be realized in solid materials. The degenerate Fermi gases can be used to explore this kind of pairing for the comparable energies between the Fermi energy, Zeeman field and SOC.


 By the tuning of the Fermi surfaces from the SOC and in-plane Zeeman field, which breaks the inversion symmetry~\cite{Zheng2014FFLO}, the FFLO state can be realized, yielding the so-called topological FFLO superfluids~\cite{gong2019PhysRevB, qunaturecom2013topological}. The Zeeman field is found to drive the topological phase transitions within the superfluid phase~\cite{PhysRevLett.103.020401,gong2011PhysRevLett,PhysRevLett.109.105302,dong2015nc,gong2015PhysRevB}. The competition between FFLO and BCS pairing near the pairing-normal phase transition has been studied within mean field theory~\cite{Fulde,YuanFu}, while results deep in the pairing phase at low temperatures are still needed. In the presence of SOC, the transition from the FFLO to BCS phase becomes continuous in one dimensional systems~\cite{Liang2015SR, roy2021}. However, the finite temperature phase transition only occurs in high dimensions, where no concrete evidence of the continuous behavior has been found.


In this manuscript, we extend the mean-field study of the superfluid phase to the numerically exact quantum Monte Carlo~(QMC) method~\cite{becca2017quantum}, in which the pairing correlation is calculated and the topological invariant in terms of quantized $\pi$ Berry phase on the spin-non-degenerate Fermi surface is applied. The spin-orbit coupled lattice model with attractive interactions have been investigated using the auxiliary field QMC method~\cite{Tang2014Berezinskii, zhangPhysRevLett2016, zhangPhysRevLett2017,ROSENBERG2019161}, in which no infamous sign problem happens in the spin-balanced case~\cite{Wu_2005}. These previous studies show that SOC can optimize the superconducting temperature, leads to rich pairing structure like the mixture of spin-singlet and spin-triplet~\cite{Tang2014Berezinskii, zhangPhysRevLett2016}, suppresses charge density wave order near the half-filling, and even contributes to the topological signature of edge current~\cite{zhangPhysRevLett2017}. However, when we include the Zeeman field, the sign problem restricts the range of parameter that we can explore, but we can still study FFLO by selecting parameters with acceptable sign~\cite{Wolak2012}.

Our results show that the ratio of Zeeman field $h$ and SOC strength $\lambda$ play an important role in determining the pairing properties, from the FFLO pairing in the small $\lambda/h$ limit to the BCS pairing in the large $\lambda/h$ limit. We present a strong evidence that the FFLO+BCS phase is stabilized between two limits. In large SOC regime, we characterize the topological properties of the superfluids by spin polarization at time-reversal-invariant momenta. We investigate the phase diagram and find the critical exponent associated with the scaling is different from the result given by the mean-field study. The topological superfluids have not yet been realized in experiments. Our results provide strong evidences for the realization of topological superfluids with the current available interactions in experiments, thus pave the way for the realization of these exotic superfluids with degenerate Fermi gases using the state-of-art setups.

\section*{Model}

We start with the Rashba spin-orbit coupled Fermi gas loaded on a 2D square optical lattice which can be described by the following fermionic Hubbard Hamiltonian:
\begin{eqnarray}
H&=&-t\sum_{\langle{\bm{i},\bm{j}}\rangle}c^{\dagger}_{\bm{i}s}c_{\bm{j}s}+ i\lambda\sum_{\langle{\bm{i},\bm{j}}\rangle} c_{\bm{i}s}^{\dagger}(\bm{\hat{e}}_{\bm{i},\bm{j}} \times \boldsymbol{\sigma})^{s,s'}_{z}c_{\bm{j}s'}\nonumber-U\sum_{\bm{i}}n_{\bm{i}\uparrow}n_{\bm{i}\downarrow}-\mu\sum_{\bm{i}}n_{\bm{i}}+ h  \sum_{\bm{i}}c^{\dagger}_{\bm{i}s} \sigma^{s,s'}_{z} c_{\bm{i}s'},
\label{hamiltonian}
\end{eqnarray}
where $c^{\dagger}_{\bm{i}s}$ and $c_{\bm{i}s}$ are the creation  and annihilation operators for fermions at site $\bm{r_i}$ with spin $s = (\uparrow,\downarrow)$. $n_{\bm{i}}=\sum_{s} n_{\bm{i}s}$ is the total density operator in real space on site $\bm{r_i}$ with $n_{\bm{i}s}=c^{\dagger}_{\bm{i}s}c_{\bm{i}s}$. The hopping amplitude between the nearest-neighbor sites $\langle{\bm{i},\bm{j}}\rangle$ is $t$, which we take equal to unity to set the energy scale. $\bm{\hat{e}}_{\bm{i},\bm{j}}$ is the unit vector connecting sites $\bm{r_i}$ and $\bm{r_j}$. $\bm{\sigma}=(\sigma_x,\sigma_y,\sigma_z)$ are the Pauli matrices.  $\lambda$ and $U(U>0)$ stand for Rashba SOC strength and  the strength of on-site attractive interaction. $\mu$ and $h$ are the chemical potential and Zeeman field, respectively. The desired filling factor $n = \sum_{\bm{i}s}\langle n_{\bm{i}s} \rangle/N$ on an $N$-site periodic supercell lattice is obtained by tuning the chemical potential $\mu$, where $\langle\dots\rangle$ is the thermodynamic average as defined in Eq.~(\ref{Eq:DQMC}) in Method section. Without interaction, the dispersion relation of square lattice model is $\epsilon_k = -2t (\cos{k_x}+\cos{k_y})-\mu$.  At low filling, the model shows similar dispersion relation to the Fermi gas with $\epsilon_k=k^2/2m-\mu'$ with $\mu'=\mu+2t$ and $m=1/2t$. We study the regime away from half filling to detect the unconventional pairing. In the current experiments of Fermi gas in the 2D optical lattices \cite{Mazurenko2017A, Gall2020Competing}, $t \sim 0.2 - 1$ kHz using $U \sim (7 - 8)t$; and the coldest temperature achieved is about $k_B T \sim 150 - 250$ Hz ($k_B$ is the Boltzman constant), which corresponds to $\beta t \sim 4$. Thus in our manuscript, we focus on the regime with $\beta t \sim 4 - 12$. 

\section*{Results}
\subsection*{Competition between FFLO pairing and BCS pairing}

\begin{figure}[hbt!]
\centering
\includegraphics[width=0.6\textwidth]{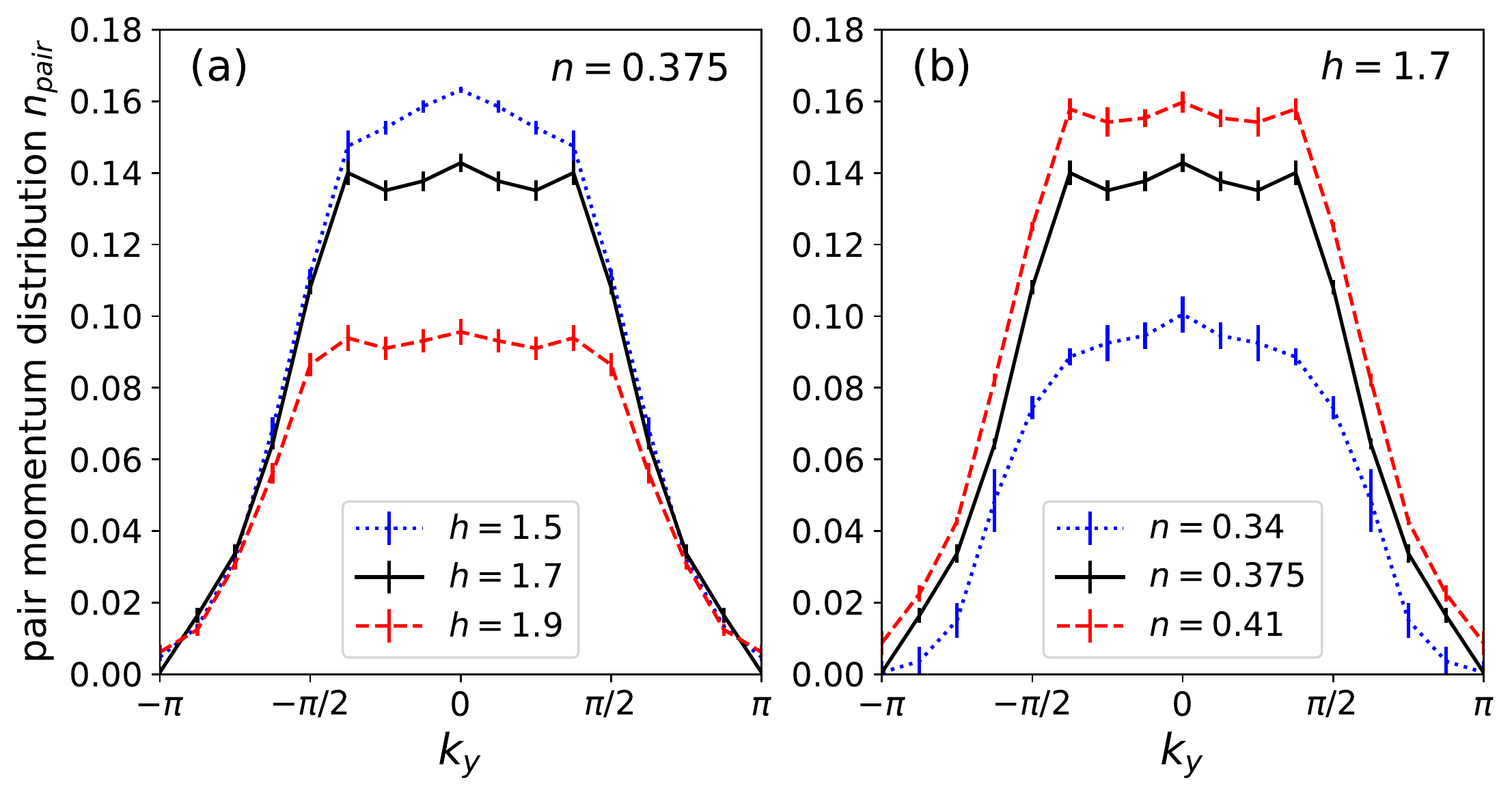}
\caption{\textbf{The FFLO+BCS pairing response to the change of the Zeeman field $h$ and the filling $n$.} (keeping $\lambda$ fixed at 0.18 as in Fig.~\ref{fig.1}(b)) We plot the pair momentum distribution $n_{pair}$ along the $k_y$ direction with $k_{x}=0$. (a) With decreasing $h$ from $1.7$ to $1.5$, the ratio of $h/\lambda$ drops, the BCS pairing peaked at $k=(0,0)$ appears~(dotted blue line). (b) Keeping the ratio of $h/\lambda$ the same, FFLO+BCS pairing persists for increasing $n$ from 0.375 to 0.41~(dashed red line). However, if the filling $n$ is small, only BCS pairing happens because of the band structure~(dotted blue line).} \label{fig.2}
\end{figure}

To investigate the effect of SOC on the pairing properties, we measure the pair momentum distribution as the Fourier transform of the pairing correlation~\cite{Liang2015SR,Feiguin2007-xd,Rizzi2008-an,Tezuka2008-xe}:
\begin{eqnarray}
n_{pair}(\bm{k})=\sum_{i,j}e^{i \bm{k}\cdot(\bm{r_i}-\bm{r_j})}\langle \Delta^{\dagger}_{\bm{i}} \Delta_{\bm{j}}\rangle,
\end{eqnarray}
where $\Delta_{\bm{i}}^{\dagger}=c_{\bm{i}\uparrow}^{\dagger} c_{\bm{i}\downarrow}^{\dagger}$ creates a singlet pair at the site $\bm{r_i}$ and $\bm{k}$ is the momentum. This distribution has a simple interpretation. For BCS pairing, it exhibits peaks at zero momentum; while for FFLO pairing between $|\bm{k}, \uparrow \rangle$ and $|-\bm{k}+\bm{Q}, \downarrow \rangle $, it exhibits peaks at $\pm \bm{Q}$.

The pair momentum distribution is shown in Fig.~\ref{fig.1} for various SOC with $U=3.5$, $h=1.7$, $\beta=10$ and filling factor $n=0.375$. In the absence of SOC, the Zeeman field lifts the degeneracy of spin. The pairing occur between different Fermi circles with different spin and hence with finite momentum. The pair momentum distribution is different from the balanced case and has a circular ridge shape. In Fig.~\ref{fig.1}(a), the peak of the distribution is at the rim of radius $|\bm{Q}|=|\bm{k}_{F\downarrow}|-|\bm{k}_{F\uparrow}|$, where $\bm{k}_{F\uparrow}$ and $\bm{k}_{F\downarrow}$ are the Fermi momenta of spin-$\uparrow$ and -$\downarrow$ Fermi circles, respectively. This nonzero momentum peak indicates the stability of the FFLO pairing in the system~\cite{wolak2012pra}.

As shown in Fig.~\ref{fig.1}(b), in the presence of SOC, the FFLO peak drops slightly. Interestingly, a zero momentum peak emerges in the distribution when the strength of SOC increases. Thus, the pair momentum distribution has a novel form: the circular ridge and the zero momentum peak coexist in the momentum space. This distribution is different from the FFLO and BCS, but contains the signatures of both two type pairing. This indicates the pairing of BCS and FFLO is compatible in a small range of SOC strength, corresponding to the FFLO-BCS pairing. In Fig.~\ref{fig.1}(c), as the SOC increases further, the zero momentum peak will ascend further until the finite circular ridge disappears in the distribution, resembling BCS pairing. We find that BCS pairing dominates for the moderate and strong strength of SOC. With increasing SOC, the pairing undergoes a transition from FFLO to BCS through FFLO-BCS.

The exotic pairing arises from the dramatic change of spin texture induced by competition between SOC and Zeeman field~\cite{Fulde,YuanFu}. The insets of Fig.~\ref{fig.1} illustrate how the spin texture in different Fermi circles changes with SOC as Zeeman field is fixed nonzero. 
Without SOC, the pairing occurs between the spin-$\uparrow$ and spin-$\downarrow$ from different bands, resulting in the finite momentum pairing of FFLO. With increasing SOC, two helical bands consist of the mixture of the spin-$\uparrow$ and spin-$\downarrow$ fermions. The FFLO pairing still occurs between two helical bands as the interband pairing, as $z$-component of the spin persists to be polarized by Zeeman field. However, BCS pairing arises from the $xy$-component of the spin within the same band as the intraband pairing. Both types of pairing coexist in the FFLO+BCS phase. Further increasing SOC, the $xy$-component of the spin becomes dominant on the Fermi surface of the helical bands, so the intraband BCS pairing dominates. 

To investigate the stability of the exotic FFLO+BCS pairing, we show how the Zeeman field and filling factor change the $n_{pair}$. In Fig.~\ref{fig.2}(a), with decreasing the Zeeman field, the BCS pairing is enhanced and the zero momentum peak becomes larger. With increasing the Zeeman field, the total pairing is suppressed, and the FFLO peak momentum becomes larger. In Fig.~\ref{fig.2}(b), when the filling factor decreases, the FFLO pairing is suppressed. For the small filling, only the lower chirality band is occupied. Only the intraband BCS pairing is allowed and the interband FFLO pairing is prohibited.



\subsection*{Topologically nontrivial superfluids and the topological phase transitions}
 
\begin{figure}[hbt!]
\centering
\includegraphics[width=0.7\textwidth]{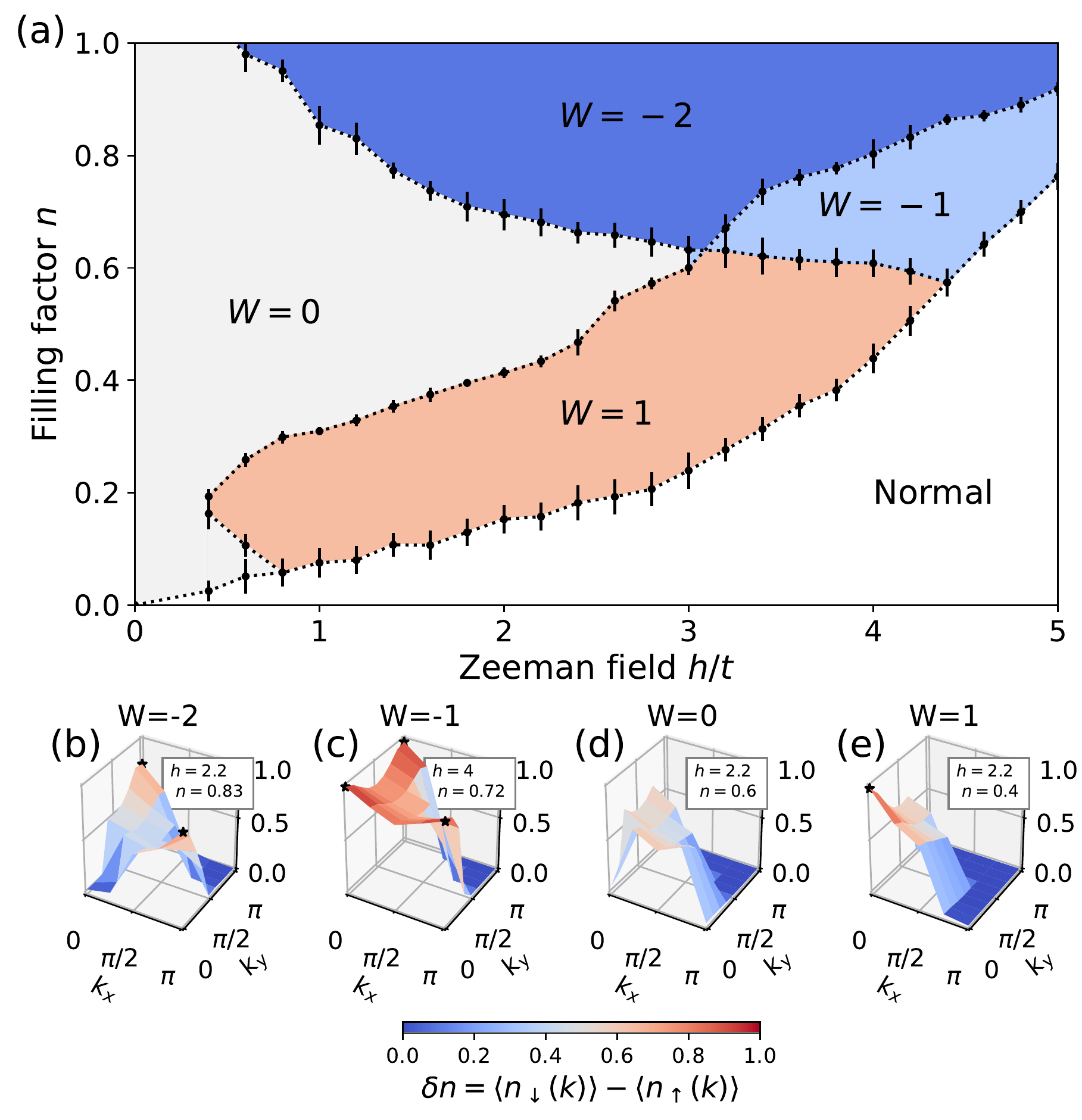}
\caption{\textbf{Phase diagram for fermions on square lattice with Rashba spin-orbit coupling and Zeeman field.} There is three topological superfluid phases with nonzero winding number $W=\pm1,-2$. Increasing the Zeeman field, there is topological phase transition from topologically trivial superfluid~($W=0$) into topologically nontrivial superfluid phases~($W\neq0$). The on-site attraction strength is $U=4$ and the spin-orbit coupling strength is $\lambda=1$. We classify the topological phase according to the spin polarization $\delta n(\textbf{k})=\langle n_{\downarrow}(\textbf{k})\rangle-\langle n_{\uparrow}(\textbf{k})\rangle$, which has jumps at time-reversal invariant momenta through the topological phase transition. The jumps are notated as ($\star$) in (b)-(e). The inverse temperature is set as $\beta=12$.} \label{fig.3}
\end{figure}

The system can realize the topological nontrivial superfluid states which break the time-reversal symmetry. The topological nontrivial phase is chiral superfluid phase, which can be characterized by winding number defined from the Green's function at all frequencies\cite{Volovik2003droplet,qixiaoliangPhysRevB2006}:
\begin{eqnarray}
W=\frac{1}{24 \pi^{2}}\int d^{2}k d\omega \mathrm{Tr}[\epsilon^{\mu\nu\rho}(\mathcal{G}\partial_{\mu}\mathcal{G}^{-1})(\mathcal{G}\partial_{\nu}\mathcal{G}^{-1})(\mathcal{G}\partial_{\rho}\mathcal{G}^{-1})],
\end{eqnarray}
where the winding number $W$ is zero for the topologically trivial phase and an nonzero integer for the topologically nontrivial phase. The winding number changes when the system undergoes a topological phase transition. The winding number can be simplified by the `topological Hamiltonian', which depends only on the zero-frequency Green's function~\cite{wangzhongPhysRevX2012,wangzhongPhysRevB2012}. But it is difficult to extract the anomalous Green's function from the QMC simulation. Thus, we do not measure the winding number directly, instead we make use of the spin polarization at the time reversal invariant (TRI) momenta~\cite{xuebingluoPhysRevA89043612,zhangnaturecomm2013topological}, since they are easy to measure in both QMC simulations and cold atom experiments. For a square lattice system, the winding number can be determined by the spin polarization at four TRI momenta $\textbf{K}=(0,0),(0,\pi),(\pi,0),(\pi,\pi)$ as follows 
\begin{eqnarray}
W&=&\sum_{\textbf{K}} \cos(K_{x}) \cos(K_{y}) \delta n(\textbf{K}),
\end{eqnarray}
where $\delta n(\textbf{K}) = n_{\downarrow}(\textbf{K})-n_{\uparrow}(\textbf{K})$ is the spin polarization at the TRI momentum $\textbf{K}$.  Prior to these results, we have verified that without SOC (setting
$\lambda = 0$), the jump of spin difference at the TRI momenta is impossible, as expected for the standard BCS superfluids
with $W = 0$. Meanwhile, without Zeeman field~(see Fig. \ref{fig.3} (a) at $h = 0$), the jump of spin difference is also
impossible, yielding a trivial superfluid with $W = 0$. There are also several special conditions: (i) The system is always
in a normal phase without cold atoms ($n=0$); (ii) This model possesses particle-hole symmetry, thus the physics of $n$ and
$(2-n)$ yields the same phase diagram with $W \rightarrow W$; (iii) This model is also symmetric about $h \rightarrow -h$
(with $W\rightarrow -W$) and $\lambda \rightarrow -\lambda$ (with $W\rightarrow W$).

Fig.~\ref{fig.3} shows the phase diagram in the $n -h$ plane with $U=4$ and $\lambda=1$. The topological invariants of the superfluid phases are determined by the spin polarization at the four TRI momenta $\delta n(\textbf{K})$. Without the Zeeman field, the system preserves the time-reversal symmetry (TRS). Then, the superfuild phase is topologically trivial.  When we introduce the Zeeman field, the TRS is broken and the superfluid phase can be topologically nontrivial. 

For small Zeeman field, the spin polarization $\delta n(\textbf{K})$ at four TRI momenta are also zero and the superfluid phase is topologically trivial with zero winding number~($W=0$).
With increasing Zeeman field, the spin polarization at the four TRI momenta jumps from $0$ into $1$ and the superfluid becomes topologically nontrivial with nonzero topological invariant. We mark the unit spin polarization at the TRI momenta as stars in Fig.~\ref{fig.3}~(b)-(e). For large filling, the spin polarization remains unchanged at $[(0,0),(\pi,\pi)]$, while has a jump from $0$ into $1$ at $[(0,\pi),(\pi,0)]$ by increasing the Zeeman field as shown in Fig.~\ref{fig.3}~(b). Therefore, the phase is topologically nontrivial with nonzero winding number~($W=-2$). By increasing the Zeeman field further, the spin polarization jumps at momentum $(0,0)$ as shown in Fig.~\ref{fig.3}~(c). Then, the topological invariant changes from $-2$ to $-1$ at the topological phase transition boundary. For intermediate filling, the spin polarization jumps from $0$ into $1$ at $(0,0)$ and remains unchange at other TRI momenta as increasing of the Zeeman field. Thus, the winding number of the topological superfluid is $W=1$ as shown in Fig.~\ref{fig.3}~(e). For the topologically nontrivial phases, the robust topological edge states also can stabilizes at the boundary of the system. The three topological superfluid phases have distinct different behavior in momentum distribution and can be easily determined in cold atom experiments. The pairing is easily to be broken by the Zeeman field and the phase is normal state with negligible pair correlation with $\mathrm{max}[n_{pair}]<0.1$ for the statistical error of the QMC method. 

We also compare the QMC phase diagram with the phase diagram obtained from the mean-field approach~(See Supplementary Material~\cite{Supplementary}). There
are some significant differences, outlined as following. (A) In the mean-field theory, the order parameter
is destroyed when $h > 3.8$; however, in the QMC simulation, the superfluids can be survived even when $h > 5$. At half filling, a much larger Zeeman field is required to destroy the pairing. (B) At low filling~($n \sim 0$), a finite Zeeman field ($h \sim 2$) is required to destroy the pairing, while from QMC, the critical Zeeman field is proportional to the filling factor $n$. These quantitative difference may imply the limitation of the mean-field theory in the understanding of the topological superfluids in degenerate Fermi gases; and the qualitatively similarities between these two approaches shed light on the realization of the topological superfluids based on the current experimental setups.


\begin{figure}[hb!]
\centering
\includegraphics[width=0.6\textwidth]{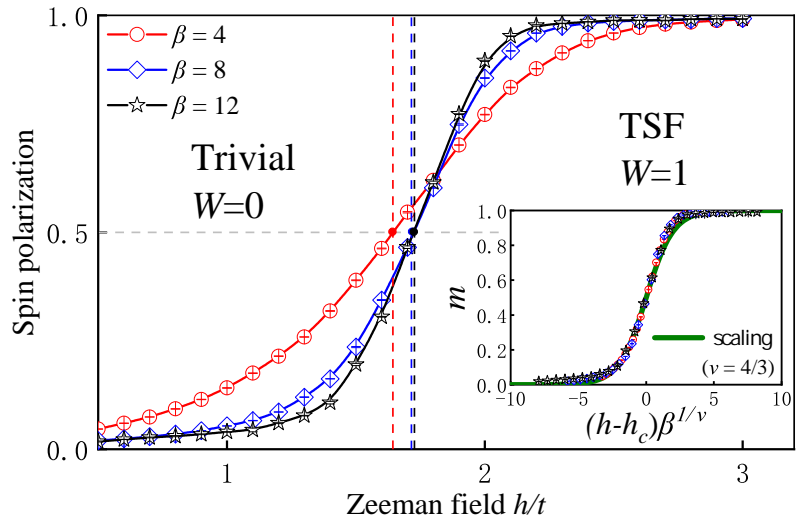}
\caption{\textbf{Topological phase transition and its associated scaling} We find the magnetization, equivalent to $\delta n(0,0)$, as a functions of Zeeman field for inverse temperature $\beta=4,8,12$ with $U=4$, $\lambda=1$ and $n=0.4$. The polarization increases from 0 into 1 as the increasing of the Zeeman field. The vertical lines with $\delta n(0,0)=0.5$ defines the phase boundary $h_c = 1.64$, $1.72$ and $1.73$ for respective temperatures. The inset shows the universal scaling law of spin polarization using Eq.~\ref{eqn4}, by which we obtain the critical exponent $\nu = 4/3 \pm 0.1$. The Green curve is the scaling function.} \label{fig.4}
\end{figure}

To probe scaling near the phase transition, we measure the spin polarization $m \equiv \delta n(0,0)$ at zero momentum as a function of Zeeman field for different temperature $\beta=1/k_{B}T=4,8,12$ with filling factor $n=0.4$ as shown in Fig.~\ref{fig.4}. The polarization $m$ approaches $1$ across the phase transition to the topologically nontrivial phase with nonzero winding number~($W=1$). In the mean field framework, the polarization is $m=\frac{1}{2}\left[ 1 + \tanh(\frac{\beta(h-h_{c})}{2})\right]$. Using the QMC result, the spin polarization as shown in insert of Fig.~\ref{fig.4}, can be  fitted by
\begin{eqnarray}
 m = \delta n(0,0)=f(\beta^{1/\nu}(h-h_c)/2)=\left[ 1 + \tanh\left(\beta^{1/\nu}(h-h_{c})/2\right)\right]/2,
 \label{eqn4}
\end{eqnarray}
where we find critical exponent $\nu=4/3\pm 0.1$. In the limit of zero temperature($\beta \rightarrow \infty$), the sharp change of polarization becomes the discontinuous jump and the critical Zeeman field is $h_c=1.74\pm0.08$.

\section*{Conclusion}

We have investigated the unconventional pairing and the topological phase transition in the spin-imbalanced spin-orbit coupled Fermi gases by unbiased QMC simulation. We study the system at an experimentally accessible temperature. We find the SOC can enrich the properties of the pairing and stabilize the exotic FFLO+BCS pairing. The interplay between FFLO pairing and BCS pairing results from the change of spin texture on the Fermi surface due to the competition between the Zeeman field and the spin-orbit coupling.  We develop the scheme to probe the topological phase transition in the QMC method by characterizing the jump of the spin polarization at time-reversal invariant momenta, in which the spin polarization can be measured in the ultracold-atom experiment directly. We find that with increasing the Zeeman field, topological phase transitions between topologically trivial superfluid phase and nontrivial phases occur, and we give the phase diagram.

The 2D spin-imbalanced spin-orbit Fermi gas could be further explored with the numerically exact QMC method. One interesting direction would be to develop the method to probe the topological signature such as edge current or other unconventional orders and to further explore the topological aspect of FFLO+BCS phase. Moreover, it would also be beneficial to explore unconventional pairing in the presence of in-plane magnetic fields, and phase diagrams of other related systems. Our results lay the theoretical foundation for the understanding of these topological superfluids, and also for the detection of these interesting physics.

\section*{Method}
Quantum Monte Carlo method is a numerically exact method
for investigating strongly correlated systems. It is especially useful in 2D, as the system size and dimensionality cause problems in other techniques like exact diagonalization and density matrix renormalization group method. In this manuscript, we use the finite temperature determinant quantum Monte-Carlo~(DQMC), whose correlation 
functions are given by
\begin{eqnarray}
 \label{Eq:DQMC}
  \langle \hat{O} \rangle &=& \frac{1}{Z} Tr \big[ \hat{O} e^{\beta H} \big] = \frac{1}{Z} \int \mathcal{D}[\phi_{i,\tau}]  
  e^{-S[\phi_{i,\tau}]} \hat{O}[\phi_{i,\tau}] \,,
\end{eqnarray}
, where $\hat{O}$ is the measurable operator, $\beta$ is the inverse temperature, and $S$ is the action with $\phi_{i,\tau}$ the set of auxiliary fields in both spatial dimension and time dimension~(please refer to literature for details~\cite{santos_2003}). 

The main step in DQMC is the matrix propagation using the propagators from interacting part $e^{-\Delta \tau H_{int}}$ and non-interacting part $e^{-\Delta \tau H_{non}}$ of the Hamiltonian, where $-\Delta \tau$ is the time interval for Trotter's decomposition. We use a Hubbard-Stratonovich transformation to convert the interaction term into a non-interacting term coupled to an discrete auxiliary field. This transformation enables us to treat Hubbard models with attractive electron interactions. If the interaction is local, $e^{-\Delta \tau H_{int}}$ is usually a diagonal matrix after the Hubbard-Stratanovich transformation. We use the discrete Hubbard-Stratanovich transformation, where the auxiliary field is of spin +1 or -1.

 To study SOC systems using DQMC~\cite{Tang2014Berezinskii,zhangPhysRevLett2016, zhangPhysRevLett2017}, we have made an advancement in the QMC methods by considering a matrix representing both spin-$\uparrow$ and spin-$\downarrow$.    Without SOC, the propagating matrix of non-interacting part is a $N\times N$ matrix, where $N$ is the number of sites. The matrix is used to do the propagation for spin-$\uparrow$ part and spin-$\downarrow$ part separately. With SOC, the size of the propagator in the simulation becomes $2N\times 2N$, as the spin-flip terms destroy the block-diagonal property in the matrix. Therefore, the computation time requires in the SOC algorithm is around 4 times more than the case without SOC, as the majority of time is spent in the matrix multiplication with time complexity of $(2N)^3$ and $N^3\times 2$ in the case of with and without SOC respectively.

Negative sign problem is an unsolved problem in the QMC method~\cite{Troyer_2005, li2019sign}. The negative weight appears in some Monte Carlo samples, leading to the exponential growth of computational time required in the case of serious sign problem. It limits the usage of QMC method to specific types of systems. In the case of spin-balanced SOC system with attractive interaction, it has been shown that no sign problem exists in the system~\cite{Wu_2005}. However, in the case of spin-imbalanced system, the sign problem arises. To handle the problem, we use the large-scale finite temperature DQMC to get the results in some regime of parameter space. In this manuscript, we carry out the simulation in $8\times16$ lattice with inverse temperature $\beta=4-12$, which corresponds to the experimental setting and ensures the convergence of our result to the ground state. Note that the temperature $\beta = 4$ has already achieved in experiments in the lattice model several years ago~\cite{Mazurenko2017A}.

\section*{Acknowledgements}
H.-Q.L. acknowledges the financial support from NSAF (No. U1930402) and NSFC (No. 11734002). X.Y. and B.H. acknowledge the support from NSFC (No.11504143, No.11547047). M.G. acknowledges the financial support from the NYTTP (No. KJ2030000001) and the NNSFC (No.GG2470000101). 

\section*{Author contributions statement}
X.Y. and H.K.T. contributed equally to this work.  X.Y., H.K.T., M.G. and H.Q.L. conceived the project. X.Y. and H.K.T. conducted the calculations. All authors took part in analyzing the results and wrote the manuscript.

\section*{Data Availability}
The data and code supporting the findings of this study are available from the corresponding authors upon request.

\bibliography{ms}

\begin{thebibliography}{10}
\urlstyle{rm}
\expandafter\ifx\csname url\endcsname\relax
  \def\url#1{\texttt{#1}}\fi
\expandafter\ifx\csname urlprefix\endcsname\relax\def\urlprefix{URL }\fi
\expandafter\ifx\csname doiprefix\endcsname\relax\def\doiprefix{DOI: }\fi
\providecommand{\bibinfo}[2]{#2}
\providecommand{\eprint}[2][]{\url{#2}}

\bibitem{RevModPhys.82.3045}
\bibinfo{author}{Hasan, M.~Z.} \& \bibinfo{author}{Kane, C.~L.}
\newblock \bibinfo{journal}{\bibinfo{title}{\textit{Colloquium} : Topological
  insulators}}.
\newblock {\emph{\JournalTitle{Rev. Mod. Phys.}}}
  \textbf{\bibinfo{volume}{82}}, \bibinfo{pages}{3045--3067},
  \doiprefix\url{10.1103/RevModPhys.82.3045} (\bibinfo{year}{2010}).

\bibitem{qiRevModPhys2011}
\bibinfo{author}{Qi, X.-L.} \& \bibinfo{author}{Zhang, S.-C.}
\newblock \bibinfo{journal}{\bibinfo{title}{Topological insulators and
  superconductors}}.
\newblock {\emph{\JournalTitle{Rev. Mod. Phys.}}}
  \textbf{\bibinfo{volume}{83}}, \bibinfo{pages}{1057--1110},
  \doiprefix\url{10.1103/RevModPhys.83.1057} (\bibinfo{year}{2011}).

\bibitem{Sato_2017}
\bibinfo{author}{Sato, M.} \& \bibinfo{author}{Ando, Y.}
\newblock \bibinfo{journal}{\bibinfo{title}{Topological superconductors: a
  review}}.
\newblock {\emph{\JournalTitle{Reports on Progress in Physics}}}
  \textbf{\bibinfo{volume}{80}}, \bibinfo{pages}{076501},
  \doiprefix\url{10.1088/1361-6633/aa6ac7} (\bibinfo{year}{2017}).

\bibitem{qixiaoliangPhysRevB2006}
\bibinfo{author}{Qi, X.-L.}, \bibinfo{author}{Wu, Y.-S.} \&
  \bibinfo{author}{Zhang, S.-C.}
\newblock \bibinfo{journal}{\bibinfo{title}{Topological quantization of the
  spin hall effect in two-dimensional paramagnetic semiconductors}}.
\newblock {\emph{\JournalTitle{Phys. Rev. B}}} \textbf{\bibinfo{volume}{74}},
  \bibinfo{pages}{085308}, \doiprefix\url{10.1103/PhysRevB.74.085308}
  (\bibinfo{year}{2006}).

\bibitem{PhysRevLett.97.240401}
\bibinfo{author}{Zhu, S.-L.}, \bibinfo{author}{Fu, H.}, \bibinfo{author}{Wu,
  C.-J.}, \bibinfo{author}{Zhang, S.-C.} \& \bibinfo{author}{Duan, L.-M.}
\newblock \bibinfo{journal}{\bibinfo{title}{Spin hall effects for cold atoms in
  a light-induced gauge potential}}.
\newblock {\emph{\JournalTitle{Phys. Rev. Lett.}}}
  \textbf{\bibinfo{volume}{97}}, \bibinfo{pages}{240401},
  \doiprefix\url{10.1103/PhysRevLett.97.240401} (\bibinfo{year}{2006}).

\bibitem{BurkovPhysRevLett2011}
\bibinfo{author}{Burkov, A.~A.} \& \bibinfo{author}{Balents, L.}
\newblock \bibinfo{journal}{\bibinfo{title}{Weyl semimetal in a topological
  insulator multilayer}}.
\newblock {\emph{\JournalTitle{Phys. Rev. Lett.}}}
  \textbf{\bibinfo{volume}{107}}, \bibinfo{pages}{127205},
  \doiprefix\url{10.1103/PhysRevLett.107.127205} (\bibinfo{year}{2011}).

\bibitem{wenhongmingPhysRevX2015}
\bibinfo{author}{Weng, H.}, \bibinfo{author}{Fang, C.}, \bibinfo{author}{Fang,
  Z.}, \bibinfo{author}{Bernevig, B.~A.} \& \bibinfo{author}{Dai, X.}
\newblock \bibinfo{journal}{\bibinfo{title}{Weyl semimetal phase in
  noncentrosymmetric transition-metal monophosphides}}.
\newblock {\emph{\JournalTitle{Phys. Rev. X}}} \textbf{\bibinfo{volume}{5}},
  \bibinfo{pages}{011029}, \doiprefix\url{10.1103/PhysRevX.5.011029}
  (\bibinfo{year}{2015}).

\bibitem{ArmitageRevModPhys2018}
\bibinfo{author}{Armitage, N.~P.}, \bibinfo{author}{Mele, E.~J.} \&
  \bibinfo{author}{Vishwanath, A.}
\newblock \bibinfo{journal}{\bibinfo{title}{Weyl and dirac semimetals in
  three-dimensional solids}}.
\newblock {\emph{\JournalTitle{Rev. Mod. Phys.}}}
  \textbf{\bibinfo{volume}{90}}, \bibinfo{pages}{015001},
  \doiprefix\url{10.1103/RevModPhys.90.015001} (\bibinfo{year}{2018}).

\bibitem{lindnernaturephy2011}
\bibinfo{author}{Lindner, N.~H.}, \bibinfo{author}{Refael, G.} \&
  \bibinfo{author}{Galitski, V.}
\newblock \bibinfo{journal}{\bibinfo{title}{Floquet topological insulator in
  semiconductor quantum wells}}.
\newblock {\emph{\JournalTitle{Nature Physics}}} \textbf{\bibinfo{volume}{7}},
  \bibinfo{pages}{490}, \doiprefix\url{10.1038/nphys1926}
  (\bibinfo{year}{2011}).

\bibitem{rudnerPhysRevX2013}
\bibinfo{author}{Rudner, M.~S.}, \bibinfo{author}{Lindner, N.~H.},
  \bibinfo{author}{Berg, E.} \& \bibinfo{author}{Levin, M.}
\newblock \bibinfo{journal}{\bibinfo{title}{Anomalous edge states and the
  bulk-edge correspondence for periodically driven two-dimensional systems}}.
\newblock {\emph{\JournalTitle{Phys. Rev. X}}} \textbf{\bibinfo{volume}{3}},
  \bibinfo{pages}{031005}, \doiprefix\url{10.1103/PhysRevX.3.031005}
  (\bibinfo{year}{2013}).

\bibitem{xuscience2015}
\bibinfo{author}{Xu, S.-Y.} \emph{et~al.}
\newblock \bibinfo{journal}{\bibinfo{title}{Discovery of a weyl fermion
  semimetal and topological fermi arcs}}.
\newblock {\emph{\JournalTitle{Science}}} \textbf{\bibinfo{volume}{349}},
  \bibinfo{pages}{613--617}, \doiprefix\url{10.1126/science.aaa9297}
  (\bibinfo{year}{2015}).

\bibitem{PhysRevLett.109.095301}
\bibinfo{author}{Wang, P.} \emph{et~al.}
\newblock \bibinfo{journal}{\bibinfo{title}{Spin-orbit coupled degenerate fermi
  gases}}.
\newblock {\emph{\JournalTitle{Phys. Rev. Lett.}}}
  \textbf{\bibinfo{volume}{109}}, \bibinfo{pages}{095301},
  \doiprefix\url{10.1103/PhysRevLett.109.095301} (\bibinfo{year}{2012}).

\bibitem{PhysRevLett.109.095302}
\bibinfo{author}{Cheuk, L.~W.} \emph{et~al.}
\newblock \bibinfo{journal}{\bibinfo{title}{Spin-injection spectroscopy of a
  spin-orbit coupled fermi gas}}.
\newblock {\emph{\JournalTitle{Phys. Rev. Lett.}}}
  \textbf{\bibinfo{volume}{109}}, \bibinfo{pages}{095302},
  \doiprefix\url{10.1103/PhysRevLett.109.095302} (\bibinfo{year}{2012}).

\bibitem{PhysRevLett.102.130401}
\bibinfo{author}{Lin, Y.-J.} \emph{et~al.}
\newblock \bibinfo{journal}{\bibinfo{title}{Bose-einstein condensate in a
  uniform light-induced vector potential}}.
\newblock {\emph{\JournalTitle{Phys. Rev. Lett.}}}
  \textbf{\bibinfo{volume}{102}}, \bibinfo{pages}{130401},
  \doiprefix\url{10.1103/PhysRevLett.102.130401} (\bibinfo{year}{2009}).

\bibitem{naturelin2009synthetic}
\bibinfo{author}{Lin, Y.-J.}, \bibinfo{author}{Compton, R.~L.},
  \bibinfo{author}{Jimenez-Garcia, K.}, \bibinfo{author}{Porto, J.~V.} \&
  \bibinfo{author}{Spielman, I.~B.}
\newblock \bibinfo{journal}{\bibinfo{title}{Synthetic magnetic fields for
  ultracold neutral atoms}}.
\newblock {\emph{\JournalTitle{Nature}}} \textbf{\bibinfo{volume}{462}},
  \bibinfo{pages}{628--632}, \doiprefix\url{10.1038/nature08609}
  (\bibinfo{year}{2009}).

\bibitem{PhysRevLett.103.035301}
\bibinfo{author}{Goldman, N.} \emph{et~al.}
\newblock \bibinfo{journal}{\bibinfo{title}{Non-abelian optical lattices:
  Anomalous quantum hall effect and dirac fermions}}.
\newblock {\emph{\JournalTitle{Phys. Rev. Lett.}}}
  \textbf{\bibinfo{volume}{103}}, \bibinfo{pages}{035301},
  \doiprefix\url{10.1103/PhysRevLett.103.035301} (\bibinfo{year}{2009}).

\bibitem{PhysRevA.84.031608}
\bibinfo{author}{Yi, W.} \& \bibinfo{author}{Guo, G.-C.}
\newblock \bibinfo{journal}{\bibinfo{title}{Phase separation in a polarized
  fermi gas with spin-orbit coupling}}.
\newblock {\emph{\JournalTitle{Phys. Rev. A}}} \textbf{\bibinfo{volume}{84}},
  \bibinfo{pages}{031608}, \doiprefix\url{10.1103/PhysRevA.84.031608}
  (\bibinfo{year}{2011}).

\bibitem{PhysRevA.87.063610}
\bibinfo{author}{Ozawa, T.}, \bibinfo{author}{Pitaevskii, L.~P.} \&
  \bibinfo{author}{Stringari, S.}
\newblock \bibinfo{journal}{\bibinfo{title}{Supercurrent and dynamical
  instability of spin-orbit-coupled ultracold bose gases}}.
\newblock {\emph{\JournalTitle{Phys. Rev. A}}} \textbf{\bibinfo{volume}{87}},
  \bibinfo{pages}{063610}, \doiprefix\url{10.1103/PhysRevA.87.063610}
  (\bibinfo{year}{2013}).

\bibitem{banerjeenaturephy2013ferromagnetic}
\bibinfo{author}{Banerjee, S.}, \bibinfo{author}{Erten, O.} \&
  \bibinfo{author}{Randeria, M.}
\newblock \bibinfo{journal}{\bibinfo{title}{Ferromagnetic exchange, spin-orbit
  coupling and spiral magnetism at the laalo\_3/srtio\_3 interface}}.
\newblock {\emph{\JournalTitle{Nat. Phys.}}} \textbf{\bibinfo{volume}{9}},
  \bibinfo{pages}{626--630}, \doiprefix\url{10.1038/nphys2702}
  (\bibinfo{year}{2013}).

\bibitem{bertnaturephysics2011direct}
\bibinfo{author}{Bert, J.~A.} \emph{et~al.}
\newblock \bibinfo{journal}{\bibinfo{title}{Direct imaging of the coexistence
  of ferromagnetism and superconductivity at the laalo3/srtio3 interface}}.
\newblock {\emph{\JournalTitle{Nat. Phys.}}} \textbf{\bibinfo{volume}{7}},
  \bibinfo{pages}{767--771}, \doiprefix\url{10.1038/nphys2079}
  (\bibinfo{year}{2011}).

\bibitem{PhysRevLett.109.085302}
\bibinfo{author}{Cole, W.~S.}, \bibinfo{author}{Zhang, S.},
  \bibinfo{author}{Paramekanti, A.} \& \bibinfo{author}{Trivedi, N.}
\newblock \bibinfo{journal}{\bibinfo{title}{Bose-hubbard models with synthetic
  spin-orbit coupling: Mott insulators, spin textures, and superfluidity}}.
\newblock {\emph{\JournalTitle{Phys. Rev. Lett.}}}
  \textbf{\bibinfo{volume}{109}}, \bibinfo{pages}{085302},
  \doiprefix\url{10.1103/PhysRevLett.109.085302} (\bibinfo{year}{2012}).

\bibitem{PhysRevLett.108.225301}
\bibinfo{author}{Li, Y.}, \bibinfo{author}{Pitaevskii, L.~P.} \&
  \bibinfo{author}{Stringari, S.}
\newblock \bibinfo{journal}{\bibinfo{title}{Quantum tricriticality and phase
  transitions in spin-orbit coupled bose-einstein condensates}}.
\newblock {\emph{\JournalTitle{Phys. Rev. Lett.}}}
  \textbf{\bibinfo{volume}{108}}, \bibinfo{pages}{225301},
  \doiprefix\url{10.1103/PhysRevLett.108.225301} (\bibinfo{year}{2012}).

\bibitem{PhysRevLett.105.160403}
\bibinfo{author}{Wang, C.}, \bibinfo{author}{Gao, C.}, \bibinfo{author}{Jian,
  C.-M.} \& \bibinfo{author}{Zhai, H.}
\newblock \bibinfo{journal}{\bibinfo{title}{Spin-orbit coupled spinor
  bose-einstein condensates}}.
\newblock {\emph{\JournalTitle{Phys. Rev. Lett.}}}
  \textbf{\bibinfo{volume}{105}}, \bibinfo{pages}{160403},
  \doiprefix\url{10.1103/PhysRevLett.105.160403} (\bibinfo{year}{2010}).

\bibitem{PhysRevLett.107.195305}
\bibinfo{author}{Yu, Z.-Q.} \& \bibinfo{author}{Zhai, H.}
\newblock \bibinfo{journal}{\bibinfo{title}{Spin-orbit coupled fermi gases
  across a feshbach resonance}}.
\newblock {\emph{\JournalTitle{Phys. Rev. Lett.}}}
  \textbf{\bibinfo{volume}{107}}, \bibinfo{pages}{195305},
  \doiprefix\url{10.1103/PhysRevLett.107.195305} (\bibinfo{year}{2011}).

\bibitem{reviewHzhai}
\bibinfo{author}{Zhai, H.}
\newblock \bibinfo{journal}{\bibinfo{title}{Spin-orbit coupled quantum gases}}.
\newblock {\emph{\JournalTitle{Int. J. Mod. Phys. B}}}
  \textbf{\bibinfo{volume}{26}}, \bibinfo{pages}{1230001},
  \doiprefix\url{10.1142/S0217979212300010} (\bibinfo{year}{2012}).

\bibitem{Mitra2018}
\bibinfo{author}{Mitra, D.} \emph{et~al.}
\newblock \bibinfo{journal}{\bibinfo{title}{Quantum gas microscopy of an
  attractive {Fermi--Hubbard} system}}.
\newblock {\emph{\JournalTitle{Nat. Phys.}}} \textbf{\bibinfo{volume}{14}},
  \bibinfo{pages}{173--177}, \doiprefix\url{10.1038/nature08482}
  (\bibinfo{year}{2018}).

\bibitem{FFstate}
\bibinfo{author}{Fulde, P.} \& \bibinfo{author}{Ferrell, R.~A.}
\newblock \bibinfo{journal}{\bibinfo{title}{Superconductivity in a strong
  spin-exchange field}}.
\newblock {\emph{\JournalTitle{Phys. Rev.}}} \textbf{\bibinfo{volume}{135}},
  \bibinfo{pages}{A550--A563}, \doiprefix\url{10.1103/PhysRev.135.A550}
  (\bibinfo{year}{1964}).

\bibitem{LOstate}
\bibinfo{author}{Larkin, A.~I.} \& \bibinfo{author}{Ovchinnikov, Y.~N.}
\newblock \bibinfo{journal}{\bibinfo{title}{Inhomogeneous state of
  superconductors}}.
\newblock {\emph{\JournalTitle{Sov. Phys. JETP}}}
  \textbf{\bibinfo{volume}{20}}, \bibinfo{pages}{762} (\bibinfo{year}{1965}).

\bibitem{Zheng2014FFLO}
\bibinfo{author}{Zheng, Z.} \emph{et~al.}
\newblock \bibinfo{journal}{\bibinfo{title}{{FFLO Superfluids in 2D Spin-Orbit
  Coupled Fermi Gases}}}.
\newblock {\emph{\JournalTitle{Sci. Rep.}}} \textbf{\bibinfo{volume}{4}},
  \bibinfo{pages}{6535}, \doiprefix\url{10.1038/srep06535}
  (\bibinfo{year}{2014}).

\bibitem{gong2019PhysRevB}
\bibinfo{author}{Huang, B.}, \bibinfo{author}{Yang, X.}, \bibinfo{author}{Xu,
  N.}, \bibinfo{author}{Zhou, J.} \& \bibinfo{author}{Gong, M.}
\newblock \bibinfo{journal}{\bibinfo{title}{Dynamical instability with respect
  to finite-momentum pairing in quenched bcs superconducting phases}}.
\newblock {\emph{\JournalTitle{Phys. Rev. B}}} \textbf{\bibinfo{volume}{99}},
  \bibinfo{pages}{014517}, \doiprefix\url{10.1103/PhysRevB.99.014517}
  (\bibinfo{year}{2019}).

\bibitem{qunaturecom2013topological}
\bibinfo{author}{Qu, C.} \emph{et~al.}
\newblock \bibinfo{journal}{\bibinfo{title}{Topological superfluids with
  finite-momentum pairing and majorana fermions}}.
\newblock {\emph{\JournalTitle{Nat. Commun.}}} \textbf{\bibinfo{volume}{4}},
  \bibinfo{pages}{2710}, \doiprefix\url{10.1038/ncomms3710}
  (\bibinfo{year}{2013}).

\bibitem{PhysRevLett.103.020401}
\bibinfo{author}{Sato, M.}, \bibinfo{author}{Takahashi, Y.} \&
  \bibinfo{author}{Fujimoto, S.}
\newblock \bibinfo{journal}{\bibinfo{title}{Non-abelian topological order in
  $s$-wave superfluids of ultracold fermionic atoms}}.
\newblock {\emph{\JournalTitle{Phys. Rev. Lett.}}}
  \textbf{\bibinfo{volume}{103}}, \bibinfo{pages}{020401},
  \doiprefix\url{10.1103/PhysRevLett.103.020401} (\bibinfo{year}{2009}).

\bibitem{gong2011PhysRevLett}
\bibinfo{author}{Gong, M.}, \bibinfo{author}{Tewari, S.} \&
  \bibinfo{author}{Zhang, C.}
\newblock \bibinfo{journal}{\bibinfo{title}{Bcs-bec crossover and topological
  phase transition in 3d spin-orbit coupled degenerate fermi gases}}.
\newblock {\emph{\JournalTitle{Phys. Rev. Lett.}}}
  \textbf{\bibinfo{volume}{107}}, \bibinfo{pages}{195303},
  \doiprefix\url{10.1103/PhysRevLett.107.195303} (\bibinfo{year}{2011}).

\bibitem{PhysRevLett.109.105302}
\bibinfo{author}{Gong, M.}, \bibinfo{author}{Chen, G.}, \bibinfo{author}{Jia,
  S.} \& \bibinfo{author}{Zhang, C.}
\newblock \bibinfo{journal}{\bibinfo{title}{Searching for majorana fermions in
  2d spin-orbit coupled fermi superfluids at finite temperature}}.
\newblock {\emph{\JournalTitle{Phys. Rev. Lett.}}}
  \textbf{\bibinfo{volume}{109}}, \bibinfo{pages}{105302},
  \doiprefix\url{10.1103/PhysRevLett.109.105302} (\bibinfo{year}{2012}).

\bibitem{dong2015nc}
\bibinfo{author}{{Dong}, Y.}, \bibinfo{author}{{Dong}, L.},
  \bibinfo{author}{{Gong}, M.} \& \bibinfo{author}{{Pu}, H.}
\newblock \bibinfo{journal}{\bibinfo{title}{{Dynamical phases in quenched
  spin-orbit-coupled degenerate Fermi gas}}}.
\newblock {\emph{\JournalTitle{Nature Communications}}}
  \textbf{\bibinfo{volume}{6}}, \bibinfo{pages}{6103},
  \doiprefix\url{10.1038/ncomms7103} (\bibinfo{year}{2015}).

\bibitem{gong2015PhysRevB}
\bibinfo{author}{Huang, B.}, \bibinfo{author}{Chan, C.~F.} \&
  \bibinfo{author}{Gong, M.}
\newblock \bibinfo{journal}{\bibinfo{title}{Large chern-number topological
  superfluids in a coupled-layer system}}.
\newblock {\emph{\JournalTitle{Phys. Rev. B}}} \textbf{\bibinfo{volume}{91}},
  \bibinfo{pages}{134512}, \doiprefix\url{10.1103/PhysRevB.91.134512}
  (\bibinfo{year}{2015}).

\bibitem{Fulde}
\bibinfo{author}{Zwicknagl, G.}, \bibinfo{author}{Jahns, S.} \&
  \bibinfo{author}{Fulde, P.}
\newblock \bibinfo{journal}{\bibinfo{title}{Critical magnetic field of
  ultra-thin superconducting films and interfaces}}.
\newblock {\emph{\JournalTitle{Journal of the Physical Society of Japan}}}
  \textbf{\bibinfo{volume}{86}}, \bibinfo{pages}{083701},
  \doiprefix\url{10.7566/JPSJ.86.083701} (\bibinfo{year}{2017}).

\bibitem{YuanFu}
\bibinfo{author}{Yuan, N. F.~Q.} \& \bibinfo{author}{Fu, L.}
\newblock \bibinfo{journal}{\bibinfo{title}{Topological metals and
  finite-momentum superconductors}}.
\newblock {\emph{\JournalTitle{Proc. Natl. Acad. Sci. U.S.A.}}}
  \textbf{\bibinfo{volume}{118}}, \doiprefix\url{10.1073/pnas.2019063118}
  (\bibinfo{year}{2021}).

\bibitem{Liang2015SR}
\bibinfo{author}{Liang, J.} \emph{et~al.}
\newblock \bibinfo{journal}{\bibinfo{title}{Unconventional pairings of
  spin-orbit coupled attractive degenerate fermi gas in a one-dimensional
  optical lattice}}.
\newblock {\emph{\JournalTitle{Sci. Rep.}}} \textbf{\bibinfo{volume}{5}},
  \bibinfo{pages}{14863}, \doiprefix\url{10.1038/srep14863}
  (\bibinfo{year}{2015}).

\bibitem{roy2021}
\bibinfo{author}{Roy, M.~S.} \& \bibinfo{author}{Kumar, M.}
\newblock \bibinfo{title}{Fulde-ferrel-larkin-ovchinnikov phase in one
  dimensional fermi gas with attractive interactions and transverse spin-orbit
  coupling} (\bibinfo{year}{2021}).
\newblock \eprint{ArXiv:2108.03314}.

\bibitem{becca2017quantum}
\bibinfo{author}{Becca, F.} \& \bibinfo{author}{Sorella, S.}
\newblock \emph{\bibinfo{title}{Quantum Monte Carlo approaches for correlated
  systems}} (\bibinfo{publisher}{Cambridge University Press},
  \bibinfo{year}{2017}).

\bibitem{Tang2014Berezinskii}
\bibinfo{author}{Tang, H.~K.}, \bibinfo{author}{Yang, X.},
  \bibinfo{author}{Sun, J.} \& \bibinfo{author}{Lin, H.~Q.}
\newblock \bibinfo{journal}{\bibinfo{title}{Berezinskii-kosterlitz-thoules
  phase transition of spin-orbit coupled fermi gas in optical lattice}}.
\newblock {\emph{\JournalTitle{EPL}}} \textbf{\bibinfo{volume}{107}},
  \bibinfo{pages}{40003}, \doiprefix\url{10.1209/0295-5075/107/40003}
  (\bibinfo{year}{2014}).

\bibitem{zhangPhysRevLett2016}
\bibinfo{author}{Shi, H.}, \bibinfo{author}{Rosenberg, P.},
  \bibinfo{author}{Chiesa, S.} \& \bibinfo{author}{Zhang, S.}
\newblock \bibinfo{journal}{\bibinfo{title}{Rashba spin-orbit coupling, strong
  interactions, and the bcs-bec crossover in the ground state of the
  two-dimensional fermi gas}}.
\newblock {\emph{\JournalTitle{Phys. Rev. Lett.}}}
  \textbf{\bibinfo{volume}{117}}, \bibinfo{pages}{040401},
  \doiprefix\url{10.1103/PhysRevLett.117.040401} (\bibinfo{year}{2016}).

\bibitem{zhangPhysRevLett2017}
\bibinfo{author}{Rosenberg, P.}, \bibinfo{author}{Shi, H.} \&
  \bibinfo{author}{Zhang, S.}
\newblock \bibinfo{journal}{\bibinfo{title}{Ultracold atoms in a square lattice
  with spin-orbit coupling: Charge order, superfluidity, and topological
  signatures}}.
\newblock {\emph{\JournalTitle{Phys. Rev. Lett.}}}
  \textbf{\bibinfo{volume}{119}}, \bibinfo{pages}{265301},
  \doiprefix\url{10.1103/PhysRevLett.119.265301} (\bibinfo{year}{2017}).

\bibitem{ROSENBERG2019161}
\bibinfo{author}{Rosenberg, P.}, \bibinfo{author}{Shi, H.} \&
  \bibinfo{author}{Zhang, S.}
\newblock \bibinfo{journal}{\bibinfo{title}{Accurate computations of rashba
  spin-orbit coupling in interacting systems: From the fermi gas to real
  materials}}.
\newblock {\emph{\JournalTitle{Journal of Physics and Chemistry of Solids}}}
  \textbf{\bibinfo{volume}{128}}, \bibinfo{pages}{161 -- 168},
  \doiprefix\url{10.1016/j.jpcs.2017.12.026} (\bibinfo{year}{2019}).

\bibitem{Wu_2005}
\bibinfo{author}{Wu, C.} \& \bibinfo{author}{Zhang, S.-C.}
\newblock \bibinfo{journal}{\bibinfo{title}{Sufficient condition for absence of
  the sign problem in the fermionic quantum monte carlo algorithm}}.
\newblock {\emph{\JournalTitle{Phys. Rev. B}}} \textbf{\bibinfo{volume}{71}},
  \bibinfo{pages}{155115}, \doiprefix\url{10.1103/PhysRevB.71.155115}
  (\bibinfo{year}{2005}).

\bibitem{Wolak2012}
\bibinfo{author}{Wolak, M.~J.}, \bibinfo{author}{Gr\'emaud, B.},
  \bibinfo{author}{Scalettar, R.~T.} \& \bibinfo{author}{Batrouni, G.~G.}
\newblock \bibinfo{journal}{\bibinfo{title}{Pairing in a two-dimensional fermi
  gas with population imbalance}}.
\newblock {\emph{\JournalTitle{Phys. Rev. A}}} \textbf{\bibinfo{volume}{86}},
  \bibinfo{pages}{023630}, \doiprefix\url{10.1103/PhysRevA.86.023630}
  (\bibinfo{year}{2012}).

\bibitem{Mazurenko2017A}
\bibinfo{author}{mazurenko, A.} \emph{et~al.}
\newblock \bibinfo{journal}{\bibinfo{title}{A cold-atom fermi–hubbard
  antiferromagnet}}.
\newblock {\emph{\JournalTitle{Nature}}} \textbf{\bibinfo{volume}{545}},
  \bibinfo{pages}{462}, \doiprefix\url{10.1038/nature22362}
  (\bibinfo{year}{2017}).

\bibitem{Gall2020Competing}
\bibinfo{author}{Gell, M.}, \bibinfo{author}{Wurz1, N.},
  \bibinfo{author}{Samland, J.}, \bibinfo{author}{Chan, C.~F.} \&
  \bibinfo{author}{H\"ohl, M.}
\newblock \bibinfo{journal}{\bibinfo{title}{Competing magnetic orders in a
  bilayer hubbard model with ultracold atoms}}.
\newblock {\emph{\JournalTitle{Nature}}} \textbf{\bibinfo{volume}{589}},
  \bibinfo{pages}{40}, \doiprefix\url{10.1038/s41586-020-03058-x}
  (\bibinfo{year}{2020}).

\bibitem{Feiguin2007-xd}
\bibinfo{author}{Feiguin, A.~E.} \& \bibinfo{author}{Heidrich-Meisner, F.}
\newblock \bibinfo{journal}{\bibinfo{title}{Pairing states of a polarized fermi
  gas trapped in a one-dimensional optical lattice}}.
\newblock {\emph{\JournalTitle{Phys. Rev. B}}} \textbf{\bibinfo{volume}{76}},
  \bibinfo{pages}{220508}, \doiprefix\url{10.1103/PhysRevB.76.220508}
  (\bibinfo{year}{2007}).

\bibitem{Rizzi2008-an}
\bibinfo{author}{Rizzi, M.} \emph{et~al.}
\newblock \bibinfo{journal}{\bibinfo{title}{{Fulde-Ferrell-Larkin-Ovchinnikov}
  pairing in one-dimensional optical lattices}}.
\newblock {\emph{\JournalTitle{Phys. Rev. B}}} \textbf{\bibinfo{volume}{77}},
  \bibinfo{pages}{245105}, \doiprefix\url{10.1103/PhysRevB.77.245105}
  (\bibinfo{year}{2008}).

\bibitem{Tezuka2008-xe}
\bibinfo{author}{Tezuka, M.} \& \bibinfo{author}{Ueda, M.}
\newblock \bibinfo{journal}{\bibinfo{title}{Density-matrix renormalization
  group study of trapped imbalanced fermi condensates}}.
\newblock {\emph{\JournalTitle{Phys. Rev. Lett.}}}
  \textbf{\bibinfo{volume}{100}}, \bibinfo{pages}{110403},
  \doiprefix\url{10.1103/PhysRevLett.100.110403} (\bibinfo{year}{2008}).

\bibitem{wolak2012pra}
\bibinfo{author}{Wolak, M.~J.}, \bibinfo{author}{Gr\'emaud, B.},
  \bibinfo{author}{Scalettar, R.~T.} \& \bibinfo{author}{Batrouni, G.~G.}
\newblock \bibinfo{journal}{\bibinfo{title}{Pairing in a two-dimensional fermi
  gas with population imbalance}}.
\newblock {\emph{\JournalTitle{Phys. Rev. A}}} \textbf{\bibinfo{volume}{86}},
  \bibinfo{pages}{023630}, \doiprefix\url{10.1103/PhysRevA.86.023630}
  (\bibinfo{year}{2012}).

\bibitem{Volovik2003droplet}
\bibinfo{author}{Volovik, G.~E.}
\newblock \emph{\bibinfo{title}{The Universe in a Helium Droplet}}
  (\bibinfo{publisher}{Oxford University Press,USA}, \bibinfo{year}{2003}).

\bibitem{wangzhongPhysRevX2012}
\bibinfo{author}{Wang, Z.} \& \bibinfo{author}{Zhang, S.-C.}
\newblock \bibinfo{journal}{\bibinfo{title}{Simplified topological invariants
  for interacting insulators}}.
\newblock {\emph{\JournalTitle{Phys. Rev. X}}} \textbf{\bibinfo{volume}{2}},
  \bibinfo{pages}{031008}, \doiprefix\url{10.1103/PhysRevX.2.031008}
  (\bibinfo{year}{2012}).

\bibitem{wangzhongPhysRevB2012}
\bibinfo{author}{Wang, Z.} \& \bibinfo{author}{Zhang, S.-C.}
\newblock \bibinfo{journal}{\bibinfo{title}{Strongly correlated topological
  superconductors and topological phase transitions via green's function}}.
\newblock {\emph{\JournalTitle{Phys. Rev. B}}} \textbf{\bibinfo{volume}{86}},
  \bibinfo{pages}{165116}, \doiprefix\url{10.1103/PhysRevB.86.165116}
  (\bibinfo{year}{2012}).

\bibitem{xuebingluoPhysRevA89043612}
\bibinfo{author}{Luo, X.}, \bibinfo{author}{Zhou, K.}, \bibinfo{author}{Liu,
  W.}, \bibinfo{author}{Liang, Z.} \& \bibinfo{author}{Zhang, Z.}
\newblock \bibinfo{journal}{\bibinfo{title}{Fidelity susceptibility and
  topological phase transition of a two-dimensional spin-orbit-coupled fermi
  superfluid}}.
\newblock {\emph{\JournalTitle{Phys. Rev. A}}} \textbf{\bibinfo{volume}{89}},
  \bibinfo{pages}{043612}, \doiprefix\url{10.1103/PhysRevA.89.043612}
  (\bibinfo{year}{2014}).

\bibitem{zhangnaturecomm2013topological}
\bibinfo{author}{Zhang, W.} \& \bibinfo{author}{Yi, W.}
\newblock \bibinfo{journal}{\bibinfo{title}{Topological
  fulde--ferrell--larkin--ovchinnikov states in spin--orbit-coupled fermi
  gases}}.
\newblock {\emph{\JournalTitle{Nat. Commun.}}} \textbf{\bibinfo{volume}{4}},
  \bibinfo{pages}{2711}, \doiprefix\url{10.1038/ncomms3711}
  (\bibinfo{year}{2013}).

\bibitem{Supplementary}
\bibinfo{note}{See Supplemental Material for details of the mean field
  studies.}

\bibitem{santos_2003}
\bibinfo{author}{Santos, R. R.~d.}
\newblock \bibinfo{journal}{\bibinfo{title}{Introduction to quantum monte carlo
  simulations for fermionic systems}}.
\newblock {\emph{\JournalTitle{Brazilian Journal of Physics}}}
  \textbf{\bibinfo{volume}{33}}, \bibinfo{pages}{36--54},
  \doiprefix\url{10.1590/S0103-97332003000100003} (\bibinfo{year}{2003}).

\bibitem{Troyer_2005}
\bibinfo{author}{Troyer, M.} \& \bibinfo{author}{Wiese, U.-J.}
\newblock \bibinfo{journal}{\bibinfo{title}{Computational complexity and
  fundamental limitations to fermionic quantum monte carlo simulations}}.
\newblock {\emph{\JournalTitle{Phys. Rev. Lett.}}}
  \textbf{\bibinfo{volume}{94}}, \bibinfo{pages}{170201},
  \doiprefix\url{10.1103/PhysRevLett.94.170201} (\bibinfo{year}{2005}).

\bibitem{li2019sign}
\bibinfo{author}{Li, Z.-X.} \& \bibinfo{author}{Yao, H.}
\newblock \bibinfo{journal}{\bibinfo{title}{Sign-problem-free fermionic quantum
  monte carlo: Developments and applications}}.
\newblock {\emph{\JournalTitle{Annual Review of Condensed Matter Physics}}}
  \textbf{\bibinfo{volume}{10}}, \bibinfo{pages}{337--356},
  \doiprefix\url{10.1146/annurev-conmatphys-033117-054307}
  (\bibinfo{year}{2019}).

\end{thebibliography}

\end{document}


\maketitle
In this supplementary material, we first outline the mean-field theory of the spin-orbit coupled Fermi gas loaded on square lattice with attractive interaction, and we give the phase diagram to compare with the quantum Monte Carlo result. Then we discuss the formula of spin polarization $m$ in the mean-field theory.

\section{Mean-field theory and mean-field phase diagram}

\begin{figure}[hbt!]
\centering
\includegraphics[width=0.6\textwidth]{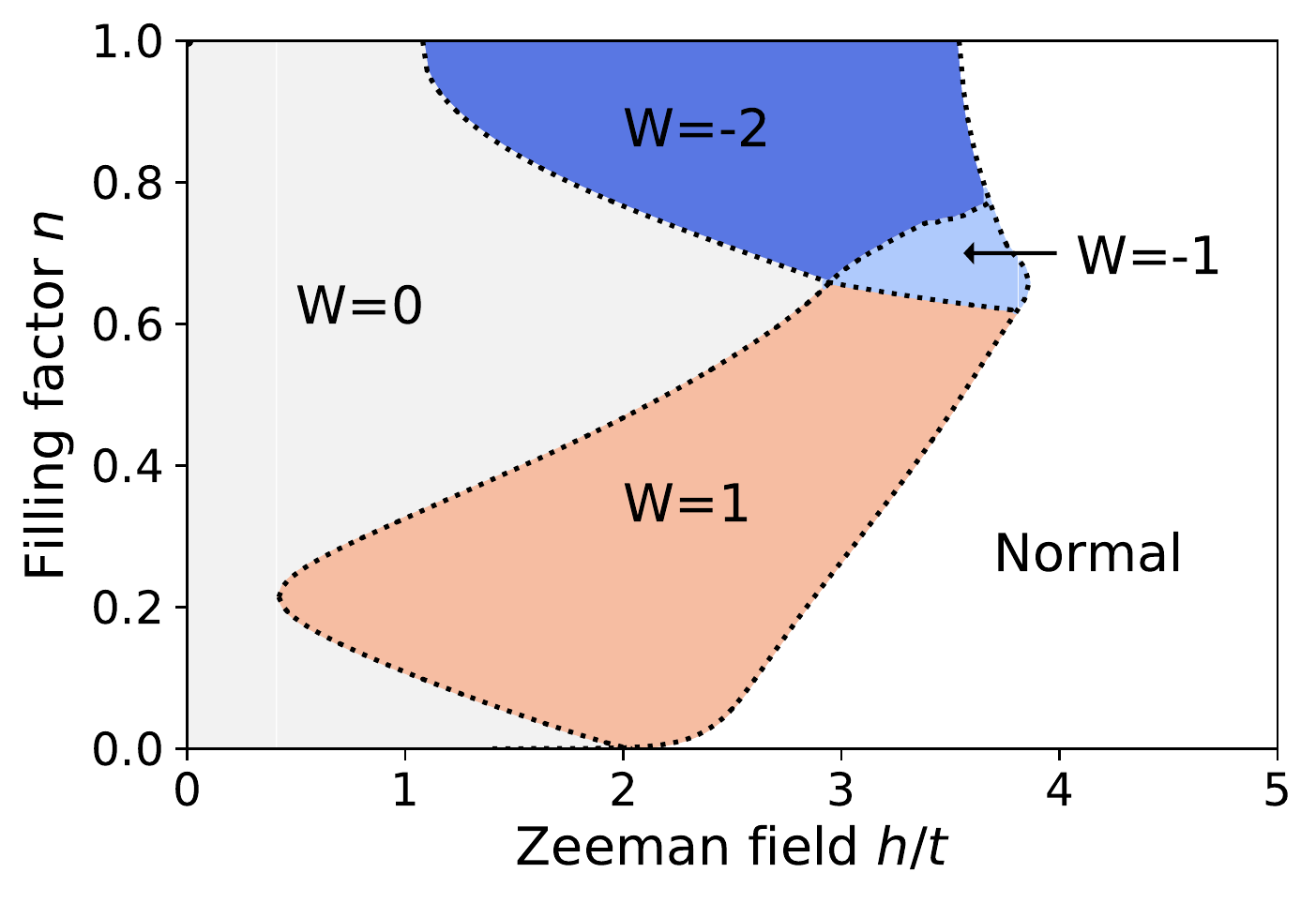}
\caption{
The mean-field phase diagram on a square lattice with Rashba SOC and Zeeman field. The on-site attraction strength is $U=4$ and the SOC strength is $\lambda=1$. The order parameter is determined self-consistently by fixing the filling factor $n$ (see Eq. \ref{eqs1} and Eq. \ref{eqs2}). The topological different phases are characterized by nonzero winding number. When the Zeeman field is strong enough, the system enters the normal phase without pairing, $\Delta = 0$.}

\label{fig.5}
\end{figure}

In mean-field framework, the strength of the BCS pairing $\Delta$ can be self-consistently determined by the gap equation with the number equation:
\begin{eqnarray}
\frac{1}{U} &=& \sum_{\mathbf{k},s=\pm}\frac{1}{4 E_{\mathbf{k},s}} \left( 1 + s \frac{h^{2}}{ E_{0}(\mathbf{k})}\right),\label{eqs1}\\
n &=& \frac{1}{2}\sum_{\mathbf{k},s=\pm} \left[1 -  \frac{\xi_{\mathbf{k}}}{E_{\mathbf{k},s}}\left(1+s \frac{h^{2}+g_{\mathbf{k}}^{2}}{E_{0}(\mathbf{k})}\right) \right].
\label{eqs2}
\end{eqnarray}
Here, $\xi_\mathbf{k}=-2t(\cos k_{x}+ \cos k_{y})-\mu$, $E_{\mathbf{k},s}=\sqrt{\xi^{2}_\mathbf{k} + h^{2} + g_{k}^{2} +\Delta^{2} +2 s E_{0}(\mathbf{k})}$, $E_{0}(\mathbf{k}) = \sqrt{h^{2}(\xi^{2}_\mathbf{k} + \Delta^{2} ) + \xi^{2}_\mathbf{k} g_{k}^{2}}$ and $\mathbf{g_{k}}=2 \lambda(\sin k_{y}, -\sin k_{x})$. The boundary between the superfluid phase and normal phase is determined by $\Delta=0$. The gap of the excitation closes at the boundaries between the topologically different superfluid phases with $h=\sqrt{(4t\pm \mu)^{2} + \Delta^{2}}$ and $h=\sqrt{\mu^{2} + \Delta^{2}}$. The superfluid is topologically trivial with $W=0$ in the absence of Zeeman field. As the Zeeman field increases, the winding number changes $1$ and $-2$ at the two boundaries, respectively. The winding number can be determined for the topologically different superfluid phases as shown in Fig.~\ref{fig.5}. The mean-field phase diagram is qualitatively consistent with the phase diagram by DQMC. There also have three topologically nontrivial superfluids
with nonzero winding numbers $W= \pm 1$, $-2$ in the phase diagram.

\section{Spin polarization in the mean-field theory}

The evolution of spin polarization in Fig. 4 in the main text, which in the mean-field framework can be derived from the polarization equation
\begin{equation}
n_{\downarrow}({\bf k})- n_{\uparrow}({\bf k}) = \frac{1}{2}\sum_{s=\pm} \left[1 + s(\frac{\xi_\mathbf{k}^{2}+\Delta^{2}}{E_{0}(\mathbf{k})})\right] \frac{h}{E_{\mathbf{k},s}} \tanh \left( \frac{\beta E_{\mathbf{k},s}}{2} \right).
\end{equation}

The gap is closed and reopened at TRI momentum
${\bf K} = (0, 0)$. At TRI momentum ${\bf K} = (0, 0)$,  the spectra $E_{\mathbf{k},s}$ defined in Eq. \ref{eqs1} and Eq. \ref{eqs2} can be rewritten as
\begin{equation}
E_{{\bf K}, \pm}= |\sqrt{(4t+\mu)^2 + \Delta^2} \pm h|.
\end{equation}
Therefore, near the critical point, we have
\begin{equation}
	E_{{\bf K},\pm} = |h \pm h_c|, \quad h_c = \sqrt{(4t+\mu)^2 + \Delta^2}.
\end{equation}
Then, the spin polarization can be obatined
\begin{equation}
	\delta n(0, 0) = {1\over 2} \left[1 + \tanh\left({1\over 2} \beta (h-h_c)\right)\right].
\end{equation}
Obviously, at the phase boundary with vanished energy gap, we have $\delta n(0, 0) =1/2$. This criteria is
used to determine the phase boundaries in our simulation,  as shown by the three vertical lines in Fig. 4 in the main text.